\documentclass{iau}
\usepackage{graphicx}
\usepackage{float}
\usepackage{color}

\newcommand{\Mjup}{\mbox{$M_{\rm Jup}$}}

\newcommand{\ie}{i.e.}

\title[Young Stars: Five Takeaways and Five Predictions] 
{Young Stars \& Planets Near the Sun in 2015:\\Five Takeaways and Five Predictions}

\author[Liu]   
{Michael C. Liu$^1$}

\affiliation{$^1$Institute for Astronomy, University of Hawaii, 2680 Woodlawn Drive, Honolulu, HI 96822, USA}

\pubyear{2015}
\volume{314}  
\pagerange{119--126}
\jname{Young Stars \& Planets Near the Sun}
\editors{J. H. Kastner, B. Stelzer, \& S. A. Metchev, eds.}

\begin{document}
\maketitle


\begin{abstract}
I present a highly biased and skewed summary of IAU Symposium 314, ``Young Stars and Planets Near the Sun," held in May 2015.  This summary includes some takeaway thoughts about the rapidly evolving state of the field, as well as some crowd-sourced predictions for progress over the next $\sim$10~years.  We predict the elimination of 1--2~of the currently recognized young moving groups, the addition of 3 or more new moving groups within 100~pc, the continued lack of a predictive theory of stellar mass, robust measurements of the gas and dust content of circumstellar disks, and an ongoing struggle to achieve a consensus definition for a planet.
\end{abstract}

\firstsection 
\section{Five Takeaways from 2015}

\begin{center}
\medskip
{\em ``Every great cause begins as a movement, becomes a business, and\\ eventually degenerates into a racket."} --- Eric Hoffer
\medskip
\end{center}

\noindent IAU Symposium~314 has been an incredible testament to the progress made in studying young stars near Earth since the last dedicated meeting on this topic almost 15 years ago, in 2001. Thanks the diligence of conference participants Matthew Kenworthy and Rahul Patel, short summaries of every conference talk are available on Facebook ({\tt \small https://goo.gl/pvHUHd}).  So like many other things in today's world, the traditional role of a conference summary talk has been superseded thanks to social media.  Thus instead of trying to recapitulate the content of this week-long meeting, instead I provide here some personal impressions about the state of the field.

\begin{enumerate}

\item[1.] {\bf Learn to live with ambiguity.}  Stellar characterization is currently experiencing a renaissance.  For young stars, there has been ample progress in key topics such as stellar age-dating, assessment of moving group membership, and spectroscopic characterization.  However, we should remember that ambiguity is inherent to this field.  Many of the key physical properties that we seek may simply be inaccessible.  

The star formation process in molecular clouds shows us that most stars form in bound clusters. The dissolution of such clusters is inevitable over time, and stars are lost from these primordial groups into the field (talk by Lada).  Thus, when we find young stars and brown dwarfs in the field, we should accept that the birth history for a significant fraction of them will be unrecoverable.  In short, not every young star that you find should belong to a moving group.

As the census of moving groups has grown, so have the methods used to assign membership, with the current pinnacle being the quantitative BANYAN method (talks by Malo \& Gagne).  Nevertheless, the known groups overlap in both velocity ($UVW$) and space ($XYZ$).  Thus, spatial and kinematic information may not be sufficient, on its own, to ever assign some young stars to a specific group. Similarly, the challenges of stellar age dating can be expected to persist, rendering an inherent fuzziness in our ability to assign ages to young stars in groups and in the field.

Finally, even objects that have a large suite of high quality measurements can present a challenge when deriving physical properties.  A very timely example is the recent activity in determining ages and masses for young brown dwarfs, from the stellar/substellar boundary all the way into the planetary mass regime.  Their young ages are reflected in their spectra through gravity-dependent absorption features.  But spectral features alone are remarkably degenerate for young objects of different masses and ages --- similar spectra (and colors and magnitudes) between objects do not necessarily correspond to the similar masses and ages (talk by Allers).

\item[2.] {\bf Everything gets more complicated the closer you look.} Fields of astronomical research often have a natural progression, as the quantity and quality of data grow.  In the early phases, data are sparse and thus simple paradigms of how things work can be remarkably successful.  This is followed by a natural compulsion to deepen and expand our data, which then sweeps away much of the initial paradigm.  At this IAU meeting, we saw many examples of the growth of complexity.  For instance, there is the long-standing picture of the nearest OB association, Sco-Cen, whereby the history of its star formation is thought to have sequentially migrated across the region.  This simple concept now needs to be revisited given the more detailed maps of its stellar distribution, which show a patchwork of different ages sewn together on the sky (talk by Mamajek). More broadly, stellar activity modeling is now building an integrated theoretical framework that combines interior modeling, angular momentum evolution, mass loss, and magnetic fields (talk by Matt). And yet we only need to look at the work being done on our Sun by the solar physics community to see how far we have to go even given a plethora of observations.

\item[3.] {\bf Initial conditions matter.} Stellar evolution theory has made significant advances in modeling the range of initial conditions relevant to the earliest stages of star formation, namely magnetic fields, accretion, rotation, and photospheric spots (talks by Feiden, Baraffe, and Somers).  This is encouraging, given the challenges of modeling these physical effects.  However, such efforts also bring to light the large role that initial conditions play in the final observational properties, even many tens to hundreds of millions of years after formation.  Since initial conditions are largely inaccessible to observations, and many are fundamentally unknowable (e.g. the early accretion history of a star), we made be heading towards a situation where the failings between models and data are naturally attributed to ``initial conditions."  Thus, in the same way that historical astronomers invoked epicycles to explain the discrepancies between the observed motion of the planets and their theoretical model (namely the Earth-centric universe), we may have found our own modern epicycles in magnetic fields, accretion, and rotation.


\item[4.] {\bf You can't always get what you want.}  A prime driver for the identification of young stars near Earth is the benefit for directly imaging exoplanets and circumstellar disks.  Tremendous gains have been made in this arena, both through the identification of the best targets (\ie, the youngest, nearest stars) and the development of high-contrast adaptive optics systems to distinguish the planets from their bright host stars.  This two-pronged approach has been enormously successful, yielding $\approx$10 directly imaged planets to date and with more expected to come imminently from Gemini/GPI, VLT/SPHERE, and SPHERE/HiCIAO.  However, during the same time that target identification and planet imaging capabilities have been improving, concurrent theoretical efforts in exoplanet formation and evolution have revealed the intense degeneracies of different formation models (cold/warm/hot start) when manifested into the available observables (luminosity, age, magnitudes, temperature, etc.).  Thus, like a small child who desires many presents on Christmas but then cannot open the shrink-wrapped boxes to get to the toys, we may soon achieve our long-sought goal of having a rich census of directly imaged planets, and yet  understanding of how they form and evolve may be elusive.

\item[5.] {\bf Stop comparing models to data and start actually testing models with data.}  ``Testing models with data" is a near-ubiquitous statement in proposals seeking telescope time and funding support.  This IAU Symposium has amply shown the advances both in the theoretical models and the observational data.  And yet it also shows the significant gap in our efforts to integrate the two aspects.  Many results shown here, and many more results published in the literature, are not truly tests of the models (Figure~\ref{fig:comparison}).  For stellar/substellar evolution, our very rich data sets are paired with state-of-the-art evolutionary and atmospheric models simply by placing the two on the same plot --- this ``chi-by-eye" approach neglects the very rich information available in the numbers, distributions, and outliers of the data, as well as the more subtle features of the models.  Similarly, in determining physical parameters for young brown dwarfs and exoplanets, an oft-used technique is to derive temperatures and gravities by overlaying models onto photometric and spectroscopic data (likewise with higher-order quantities such as clouds and chemistry, as well as ``uncertainties" on all these  quantities).  This ``overlay" method is inherently an open loop process, with no actual validation of the models used to derive the physical parameters.  We should endeavor to do better in order to yield the full fruit of our efforts.

\begin{figure}[h]
\vspace*{-1 cm}
\hbox{
   \hskip -1cm 
   \includegraphics[angle=0, width=6.0in]{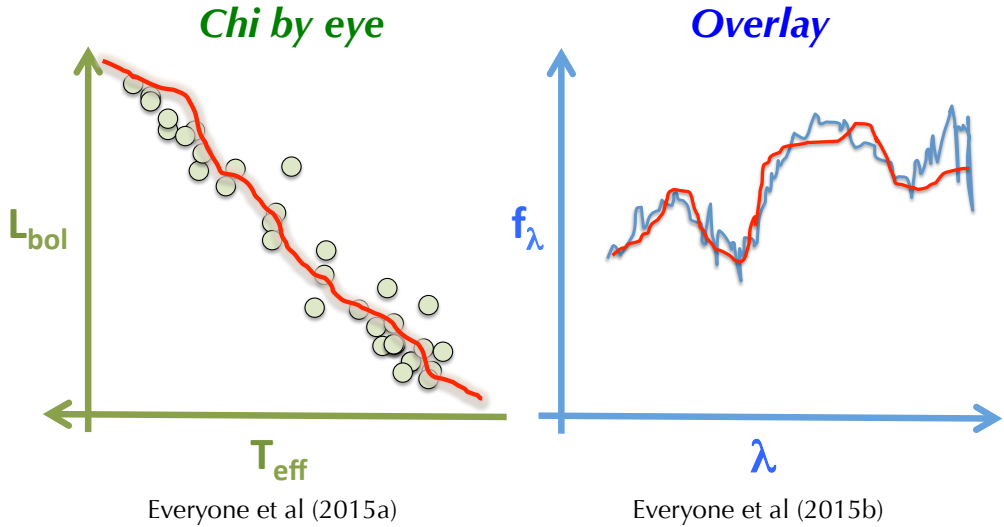} 
}
\vspace*{-3 cm}
 \caption{Representation of common practices for the interplay of models and data.  We need to do better. \label{fig:comparison}}
\end{figure}

\end{enumerate}

\section{Five Predictions for 2025}

The future is very bright for study of young stars and planets near Earth given the wide range of upcoming astronomical facilities on the horizon, e.g., ALMA, Pan-STARRS, GAIA, eROSITA, JWST, LSST, and a host of ELTs.  We can easily see the tremendous growth in this field by comparing the advances from the 2001 meeting and our 2015 gathering. What does the next decade hold?

\begin{figure}
\begin{center}
\includegraphics[angle=0, width=5.0in]{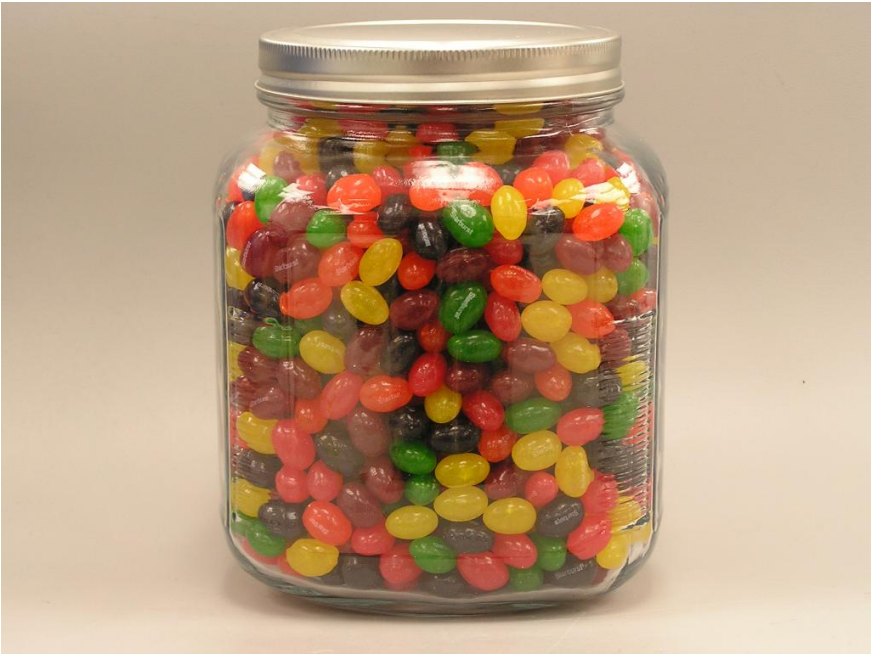} 
 \caption{The wisdom of the masses is a common method to discern the answer to questions that are too difficult for any individual to answer, such as the number of jellybeans in a jar. \label{fig:jellybeans}}
\end{center}
\end{figure}

In an attempt to forecast the future (and to provide some legacy value for this conference proceedings), the audience\footnote{This refers to the audience that was actually present for the final talk of the conference, which amounted to 68\% of the registered participants.} was presented with 5 key scientific questions that will be addressed in the next 10~years. Since prognostication is a difficult task for any one individual, we invoked the method of the ``wisdom of the masses."  A common example of this method is counting jellybeans in a jar: it is hard for any one person to guess the right number, but the median of guesses from a large group of people will yield an accurate result (Figure~\ref{fig:jellybeans}).  In this spirit, the audience voted on three possible outcomes for each of the key science questions.  The results are summarized here.  (Note that the vote-counting was done in real-time from the podium, so it may not be exactly accurate.  Not everyone voted on every question, so the total number of votes for each question is not the same.)

\begin{enumerate}

\item[1.]{\bf How many of the currently known young moving groups will be eliminated?} Some of the known YMGs are quite secure in terms of their legitimacy, while the reality of others is currently being debated.  For reference, we consider the 7~groups adopted in the BANYAN model (TWA, $\beta$~Pic, AB~Dor, Tuc-Hor, Columba, Argus \& Carina) as the current census.  {\em CHOICES: None (7~votes), 1--2 (82~votes), $>$2 (3~votes).}

\item[2.]{\bf How many new young groups will be found within 100~pc?} With the  revolutionary astrometric datasets from GAIA and LSST, as well as complementary massive spectroscopic surveys and the new all-sky X-ray mission eROSITA, we can anticipate significant progress in finding/defining new moving groups.  {\em CHOICES: None (2~votes), 1--2 (25~votes), $>$2 (65~votes).}

\item[3.]{\bf Will we have a complete predictive theory of how stars get their mass?}  While observational studies have made great advances in mapping the physical conditions and resulting distributions (in mass, age, position, etc.) of forming stars and their natal molecular clouds, a predictive theory of the stellar initial mass function remains an outstanding challenge.  {\em CHOICES: Yes (1~vote -- Chabrier), No (91~votes).}

\item[4.]{\bf Will we be able to robustly measure the gas and dust masses of circumstellar disks?} We are at the start of a great observational revolution in sub-millimeter astronomy thanks to ALMA.  To advance studies of disks and the associated planet formation process, an accurate inventory of the gas and dust components is needed. {\em CHOICES: Yes (48~votes), No (27~votes).} 

\item[5.]{\bf Will we have a consensus definition of a ``planet"?} The diversity of objects that might warrant the prized label of a planet has expanded now that observational facilities are capable of directly detecting objects down to a few Jupiter masses, both free-floating and as companions around higher mass objects.  The 2003 definition produced by the International Astronomical Union, using the deuterium-burning limit ($\approx$13~\Mjup) as the dividing line between planets and brown dwarfs, has amply been shown to be an inadequate description of nature.  Will we have a clear view on this in 10 years?  {\em CHOICES: Yes (14~votes), No (78~votes).}

\end{enumerate}

\begin{figure}[h]
\hbox{
   \hskip -1cm 
   \includegraphics[angle=0, width=5.5in]{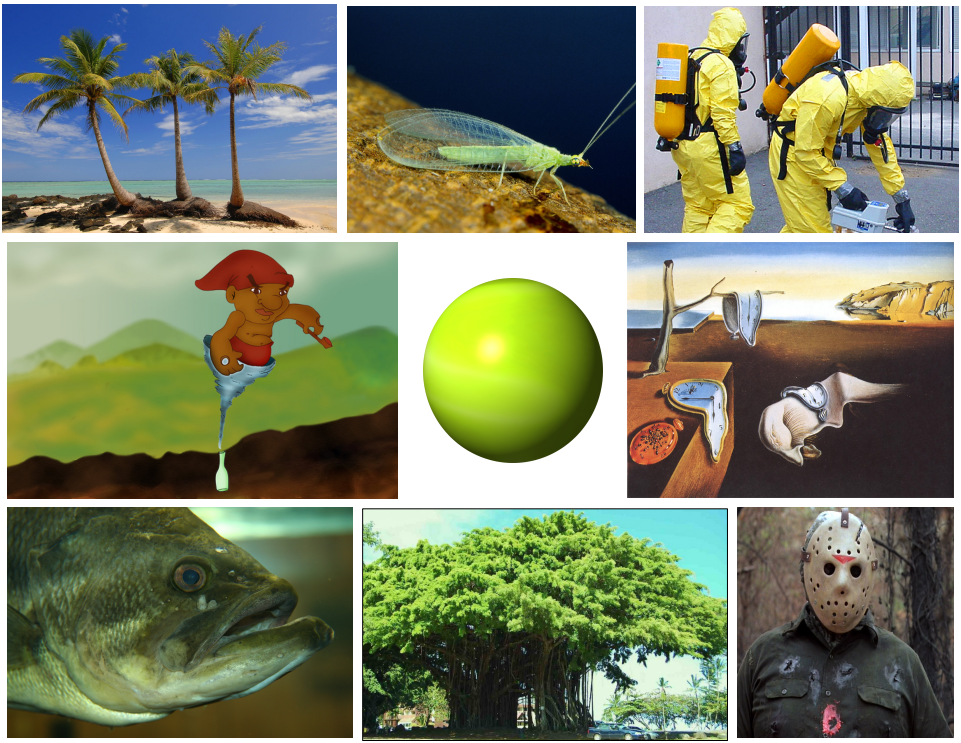} 
}
 \vspace*{-0.5 cm}
 \caption{A selection of acronyms presented at this meeting.  {\em Top row:} PALMS (B.~Bowler), LACEWING (A.~Riedel), HAZMAT (E.~Shkolnik).  {\em Middle row:} SACY (P.~Elliott), SPHERE (G.~Chauvin), DALI (N.~van der Marel). {\em Bottom row:} BASS (J.~Gagne), BANYAN (L.~Malo), JASON (S.~Murphy).}
\end{figure}

%
%

\end{document}